\documentclass[aps,twocolumn,prd,reprint,superscriptaddress,nofootinbib,floatfix]{revtex4-2}

\usepackage{amsmath}
\usepackage{amstext, amssymb, amsthm, amsfonts, slashed}

\usepackage[utf8]{inputenc} 
\usepackage{graphicx}
\usepackage{dcolumn}
\usepackage{bm}
\usepackage[bookmarks, pagebackref=false]{hyperref}
\usepackage[nameinlink]{cleveref}
\usepackage[usenames,dvipsnames]{xcolor}
\usepackage[normalem]{ulem}
\usepackage{physics}
\definecolor{rossoCP3}{cmyk}{0,.88,.77,.40}
\definecolor{blaa}{rgb}{0.2,0.2,0.6}
\hypersetup{colorlinks, 
	bookmarksopen, 
	bookmarksnumbered,
	citecolor=blaa, 		
	linkcolor=rossoCP3,	
	urlcolor=rossoCP3,			
}

\def\Tr{~{\mbox{Tr}}}

\crefname{section}{Sec.\!}{Secs.\!}
\crefname{figure}{Fig.\!}{Figs.\!}
\crefname{equation}{}{}
\crefname{table}{Tab.\!}{Tabs.\!}
\crefname{appendix}{App.\!}{Apps.\!}

\usepackage{natbib}
\usepackage{slashed}
\newcolumntype{x}[1]{>{\centering\arraybackslash\hspace{0pt}}p{#1}}

\graphicspath{{./figs/}}

\begin{document}
 
\title{ \LARGE  \color{rossoCP3} Gravitational Waves from Composite Dark Sectors}

\author{Roman {\sc Pasechnik}}
\thanks{{\scriptsize Email}: \href{mailto:roman.pasechnik@hep.lu.se}{roman.pasechnik@hep.lu.se}; \\ {\scriptsize ORCID}: \href{https://orcid.org/0000-0003-4231-0149}{0000-0003-4231-0149}}
\affiliation{Department of Physics, Lund University, 22100 Lund, Sweden}

\author{Manuel {\sc Reichert}}
\thanks{{\scriptsize Email}: \href{mailto:m.reichert@sussex.ac.uk}{m.reichert@sussex.ac.uk}; {\scriptsize ORCID}: \href{https://orcid.org/0000-0003-0736-5726}{ 0000-0003-0736-5726}}
\affiliation{Department  of  Physics  and  Astronomy,  University  of  Sussex,  Brighton,  BN1  9QH,  U.K.}

\author{Francesco {\sc Sannino}}
\thanks{{\scriptsize Email}: \href{mailto:sannino@qtc.sdu.dk}{sannino@qtc.sdu.dk}; {\scriptsize ORCID}: \href{https://orcid.org/0000-0003-2361-5326}{ 0000-0003-2361-5326}}
\affiliation{Quantum Theory Center ($\hbar$QTC), Danish-IAS, IMADA, University of Southern Denmark, Campusvej 55, 5230 Odense M, Denmark} 
\affiliation{Dipartimento di Fisica “E. Pancini”, Università di Napoli Federico II | INFN sezione di Napoli, Complesso Universitario di Monte S. Angelo Edificio 6, via Cintia, 80126 Napoli, Italy}

\author{Zhi-Wei {\sc Wang}}
\thanks{{\scriptsize Email}: \href{mailto:zhiwei.wang@uestc.edu.cn}{zhiwei.wang@uestc.edu.cn};
{\scriptsize ORCID}: \href{https://orcid.org/0000-0002-5602-6897}{0000-0002-5602-6897}} 
\affiliation{School of Physics, The University of Electronic Science and Technology of China,\\ 88 Tian-run Road, Chengdu, China}

\begin{abstract}
We study under which conditions a first-order phase transition in a composite dark sector can yield an observable stochastic gravitational-wave signal. To this end, we employ the Linear-Sigma model featuring $N_f=3,4,5$ flavours and perform a Cornwall-Jackiw-Tomboulis computation also accounting for the effects of the Polyakov loop. The model allows us to investigate the chiral phase transition in regimes that can mimic QCD-like theories incorporating in addition composite dynamics associated with the effects of confinement-deconfinement phase transition. A further benefit of this approach is that it allows to study the limit in which the effective interactions are weak. We show that strong first-order phase transitions occur for weak effective couplings of the composite sector leading to gravitational-wave signals potentially detectable at future experimental facilities.
\end{abstract} 
\maketitle

\section{Introduction}
\label{sec:introduction}

Present and future gravitational-wave (GW) detectors add extra dimensions to the way we observe and explore the world around us \cite{LIGOScientific:2016aoc, LIGOScientific:2017vwq}. The added dimension for beyond the Standard Model physics is the possibility to investigate a number of physical scenarios in a complementary way to direct observations, for example, at colliders \cite{Kosowsky:1992rz} (for a recent review, see e.g.~\cite{Athron:2023xlk}). The existing and planned GW observatories such as NANOGrav \cite{McLaughlin:2013ira, Brazier:2019mmu} and Laser Interferometer Space Antenna (LISA) \cite{LISA:2017pwj, Barausse:2020rsu} can detect the stochastic GW background sourced by violent cosmological events such as first-order phase transitions in the early universe. Such events are typically very sensitive to the presence of new particles and interactions determining the shape of the effective potential at finite temperatures. For an overview of the existing new physics realisations that feature the strongly first-order phase transitions, see for instance Ref.~\cite{Caprini:2019egz}, and for possible new physics implications of the recent NANOGrav measurement of the stochastic GW background, see e.g.~Ref.~\cite{NANOGrav:2023hvm} and references therein.

Among such possibilities, an exciting one is that our universe may feature new composite dynamics. Nussinov first considered this possibility in \cite{Nussinov:1985xr} within models of dynamical electroweak symmetry breaking in which dark matter emerges as dark baryons. It was, however, later understood that new composite asymmetric pions \cite{Gudnason:2006ug,Gudnason:2006yj} are interesting candidates for dark matter triggering  the first dedicated lattice study of composite dark matter \cite{Lewis:2011zb}. Having constructed concrete underlying extensions of the Standard Model in terms of new fundamental composite dynamics it became clear that their thermal history could be investigated via gravitational wave detection \cite{Jarvinen:2009mh}. 

More recent applications of composite dynamics to the dark sector of the universe can be found in \cite{DelNobile:2011je, Hietanen:2013fya, Bai:2013xga, Hochberg:2014dra, Pasechnik:2014ida, Antipin:2015xia, Schwaller:2015tja, Cline:2016nab, Kribs:2016cew, Dondi:2019olm, Ge:2019voa, Beylin:2019gtw, Yamanaka:2019yek, Yamanaka:2019aeq, Cacciapaglia:2020kgq, Asadi:2021yml, Asadi:2021pwo, Carenza:2022pjd, Carenza:2023shd} literature. Our most recent contributions to the field started with asking the simple question: Can the simplest model of dark composite dynamics be detectable by present and future GWs observatories \cite{Huang:2020crf,Yang:2022ghc}? Arguably, the simplest model for dark composite dynamics is the one stemming from a purely dark gluonic sector interacting only with itself and gravity. Although strong dynamics are not yet fully solved the inevitable dark confinement phase transition occurring during the universe's evolution leads to a primordial background of GWs. In our work, we combined symmetries, effective approaches and state-of-the-art lattice results to arrive at precise predictions \cite{Huang:2020crf} (see also \cite{Halverson:2020xpg, Kang:2021epo, Reichert:2022naa, Morgante:2022zvc, Yang:2022ghc, Lucini:2023irm, Bennett:2023gbe, Chen:2017cyc,Agashe:2019lhy, Bigazzi:2020avc, Garcia-Bellido:2021zgu, Ares:2020lbt, Zhu:2021vkj, Li:2021qer}). While much is still left to be understood even for such a simple model, it is a fact that composite dynamics is much richer and besides dark gluonic degrees of freedom can feature new light matter. The latter greatly alters the phase diagram of the theory adding to the confinement phase transition, typically also new ones such as the chiral phase transition \cite{Helmboldt:2019pan, Reichert:2021cvs, Yang:2022ghc}, and even a quantum one when modifying the number of matter fields to achieve a conformal phase.

It is a fact that for such a wide spectrum of composite theories, we lack a full understanding because strongly coupled dynamics is yet unsolved. Among the various methodologies we could use to tackle the dynamics the lattice field theory would be ideal but, unfortunately, lattice methods are expensive and often still too immature to investigate the full dynamics of these theories. Other methods make use of toy models of holographic nature \cite{Gursoy:2007cb, Gursoy:2007er}. 

We resort, here, to the time-honoured approach of effective theories such as the Linear Sigma Model (LSM). The reason for such a choice is two-fold. The first is that there is a great body of literature that allows us to test our results and the second is that in certain coupling regimes when the theory becomes weakly coupled, the computations allow us to extend our results also to perturbative models. The latter feature is, for example, absent in holographic approaches which by construction can only access some features of the strongly interacting composite dynamics. Specifically, in this work, we use the Polyakov quark meson model (PQM) also known as Polyakov-loop improved Linear Sigma Model (PLSM) \cite{Schaefer:2007pw, Kahara:2008yg, Schaefer:2008ax} as an effective theory to study the first-order chiral phase transition for a generic number of Dirac matter fields $N_f$. As a finite-temperature computational framework, we use the Cornwall-Jackiw-Tomboulis (known as CJT) method which allows us to go beyond the mean-field approximation \cite{Cornwall:1974vz, Amelino-Camelia:1992qfe, Amelino-Camelia:1993rvt}. As advertised above, the (P)LSM with the CJT method compared to other model computations such as holography \cite{Gursoy:2007cb, Gursoy:2007er, Ares:2021ntv, Ares:2021nap, Chen:2022cgj} and the Polyakov-loop improved Nambu-Jona-Lasinio (PNJL) model \cite{Fukushima:2003fw, Ratti:2005jh, Helmboldt:2019pan, Reichert:2021cvs}, can bridge perturbative and non-perturbative regimes of the effective theory corresponding to distinct underlying dynamical realizations.

Our main findings can be summarised as follows. Stronger first-order phase transitions appear for less strongly coupled composite models and light sigma masses expressed in units of the pion decay constant. When considering the phase transition as a function of the number of flavours we find that stronger phase transitions appear when increasing the number of flavours. However, exceptions to this behaviour occur depending on the interplay, for example, between the anomaly-induced terms and the values assumed by the effective parameters of the theory. Specifically, as the sigma mass in units of the critical temperature decreases the order of the phase transition increases for different $N_f$. As mentioned above, intriguingly for $N_f=3$ such a dependence on the lightness of the sigma state suddenly amplifies when the mass drops around twice the critical temperature in part due to the presence of a relevant anomaly-induced operator in the action.

We then translate the information about the type and strength of the studied phase transitions into potential GW-related signals. We show that these will be observable at the proposed future BBO \cite{Crowder:2005nr, Corbin:2005ny, Harry:2006fi,Thrane:2013oya, Yagi:2011wg} and DECIGO \cite{Seto:2001qf, Yagi:2011wg, Kawamura:2006up, Isoyama:2018rjb} GW experiments but fall short of the LISA expected sensitivity \cite{Audley:2017drz, Baker:2019nia, LISAdocument, LISACosmologyWorkingGroup:2022jok}. A summary of the sensitivity curves for the above measurements is given in Ref.~\cite{Schmitz:2020syl}. 

The paper is organised as follows. In \cref{sec:theory}, we present the theoretical formulation of the PLSM incorporating finite temperature effects via the CJT approach. In \cref{sec:phase-transition}, we briefly overview the standard picture of the first-order phase transition dynamics and characteristics of the resulting GW power spectrum. In \cref{sec:results}, we present the main results of our analysis including the flavour dependence, the effect of the Polyakov loop sector as well as the impact of the determinantal term. The physics implications of our results are discussed in \cref{Sec:discussion}. Finally, in \cref{sec:conclusion} we summarise our main conclusions.

\section{Theoretical setup}
\label{sec:theory}

In this section, we briefly summarize the effective actions we use to investigate the order and strength of the phase transition as well as the computational method used to extract the finite temperature corrections relevant to obtain the desired information.

\subsection{Polyakov-Loop Improved Linear Sigma Model}
\label{sec:PLSM}

The model we use combines information about confinement contained in the Polyakov loop with the physics of chiral symmetry encoded in the LSM. The resulting effective model has different names in the literature from the PQM to the PLSM. An alternative approach is the PNJL model \cite{Fukushima:2003fw, Ratti:2005jh, Schaefer:2007pw, Kahara:2008yg, Schaefer:2008ax}, which prefers to couple the Polyakov loop to the non-renormalizable NJL model. It is fair to say that the approaches are complementary and feature their own set of well-known shortcomings. 

The Lagrangian of the PLSM where mesons couple to a spatially constant temporal background gauge field $A_0$ reads
\begin{align}
 \mathcal{L}&=\bar{q}\left(i D\! \! \! \!/-g\left(\sigma+i\gamma_5T^a\pi_a\right)\right)q+\frac{1}{2}\left(\partial_\mu\sigma\right)^2+\frac{1}{2}\left(\partial_\mu\pi_a\right)^2 \notag \\
 &\quad\, -V_{\rm{PLM}}^{\mathrm{(poly)}}+V_{\text{LSM}}+V_{\rm{medium}}\,,
\end{align}
where $D\! \! \! \!/=\gamma_\mu\partial_\mu-i\gamma_0 A_0$ and $V_{\rm{PLM}}^{\mathrm{(poly)}}+V_{\text{LSM}}+V_{\rm{medium}}$ denote, respectively, the Polyakov-loop potential, the LSM potential and the interaction terms between the Polyakov loop and meson fields. All three potentials are explained in detail below.

In the pure gluon sector of a $SU(N)$ Yang-Mills theory, a centre symmetry $Z_N$ is used to distinguish the confinement and deconfinement phases. The corresponding order parameter is the colour trace of the Polyakov loop
\begin{align}	
{\ell}_F\left(\vec{x}\right)=\frac{1}{N}{\rm Tr}_c[{\bf L}_F]\,,
\nonumber
	\label{eq:Polyakov_Loop}
\end{align}
where $F$ represents fundamental representation of the gauge group, and ${\bf L}_F$ is the Polyakov loop defined as
\begin{align}
	{\bf L}_F(\vec{x})={\mathcal P}\exp\!\left[i\int_{0}^{1/T} \!\! A_{4}(\vec{x},\tau)\,\mathrm d\tau\right]\,.
\end{align}
Here, $A_4=A_4^aT^a$ and $T^a$ are the generators in the defining (fundamental) representation of the gauge group. More generically, we can define ${\bf L}_R$ similarly with generators $T_R$ in a general representation $R$. In what follows, we focus on the fundamental representation and denote, for simplicity, ${\bf L} \equiv{\bf L}_F$ and ${\ell} \equiv {\ell}_F$.

The thermal effective potential of the gluon sector which preserves the $Z_{N}$ symmetry in the polynomial form is described by the Polyakov loop as \cite{Pisarski:2000eq, Pisarski:2001pe}:
\begin{align}
	&V_{\rm{PLM}}^{\mathrm{(poly)}}=T^4\left(-\frac{b_2(T)}{2}|\ell|^2+b_4|\ell|^4+{\textcolor{red}{\cdots}}-b_3\!\left(\ell^{N}+\ell^{*N}\right)\right)\nonumber  \\ 
& b_2(T)=a_0+a_1\!\left(\frac{T_0}{T}\right)\!+a_2\!\left(\frac{T_0}{T}\right)^{\!2}\!
	+a_3\!\left(\frac{T_0}{T}\right)^{\!3}\!+a_4\!\left(\frac{T_0}{T}\right)^{\!4} \,.
	\label{eq:PLM_potential}
\end{align}
The above \textcolor{red}{``$\cdots$"} represent any required lower-dimension operator than $\ell^N$ i.e.~$\left(\ell\ell^*\right)^k=|\ell|^{2k}$ with\,\,$2k<N$. In \cite{Sannino:2002wb} it was first realized how to transfer the information about confinement encoded in the somewhat mathematical Polyakov loop model to the hadronic quasiparticle states of the theory and vice versa. The approach was later generalized to take into account an interplay between confinement and chiral symmetry breaking \cite{Mocsy:2003qw,Mocsy:2003tr}.  

For the LSM we assume the symmetry to be $SU(N_f)\times SU(N_f)$ for a given number of flavours $N_f$ and the potential reads \cite{Meurice:2017zng}
\begin{align}
V_{\text{LSM}}&=\frac{1}{2} \left(\lambda _{\sigma }-\lambda _a\right) \text{Tr}\!\left[\Phi ^{\dagger }\Phi \right]^2+\frac{N_f}{2} \lambda _a \text{Tr}\!\left[\Phi ^{\dagger }\Phi \Phi ^{\dagger }\Phi \right] \notag \\
&\quad -m^2 \text{Tr}\!\left[\Phi ^{\dagger }\Phi \right] \notag \\
&\quad -2\left(2N_f\right)^{N_f/2-2} c \left( \text{det}\,\Phi ^{\dagger } + \text{det}\,\Phi \right)\,,\label{VLSM}
\end{align}
where the meson field $\Phi$ is a $N_f\times N_f$ matrix defined as
\begin{align}
\Phi=\frac{1}{\sqrt{2N_f}}\left(\sigma+i\eta'\right)I+\left(a_a+i\pi_a\right)T^a\,,
\label{meson_field}
\end{align}
where $I$ denotes the identity matrix and $T^a$ is the corresponding $SU(N_f)$ generators with $a=1,\,2\cdots N_f^2-1$. $\Phi$ is invariant under $U(N_f)\times U(N_f)$ transformation. The determinant term in \cref{VLSM} breaks the $U(1)_A$ symmetry and thus the potential \cref{VLSM} is invariant under $SU(N_f)_L\times SU(N_f)_R\times U(1)_V$. Furthermore, the $\sigma$ field in \cref{meson_field} acquires a vacuum expectation value (VEV) triggering the spontaneous chiral symmetry breaking i.e.~$SU(N_f)_L\times SU(N_f)_R\times U(1)_V\rightarrow SU(N_f)_V\times U(1)_V.$ We have
\begin{align}
    \langle \Phi\rangle &= \frac{v}{\sqrt{2N_f}}\,,
    &
    f_\pi &=\sqrt{2/N_f}\,v\,,
    \label{eq:pion_decay_constant}
\end{align}
where $f_\pi$ is the pion decay constant. The interaction between the Polyakov loop and meson fields is described by the medium potential $V_{\rm{medium}}\!\left[\langle \bar{\psi}\psi\rangle, \ell,\ell^*\right]$ evaluated as follows~\cite{Fukushima:2017csk}
\begin{align}
	\label{eq:medium-pot}
	V_{\rm{medium}}\!\left[\langle \bar{\psi}\psi\rangle, \ell,\ell^*\right]
	=-2N_f T\int\!\! \frac{\mathrm d^3p}{(2\pi)^3}(S_R+S_R^\dagger)\,,
\end{align}
in which $S_R$ and $S_R^\dagger$ are defined for a representation $R$ as
\begin{align}
	S_R &\equiv\Tr_C\ln\!\left[1+{\bf L}_R\exp\!\Big(-\frac{E_p-\mu}{T}\Big)\right], \notag \\
	S_R^\dagger &\equiv\Tr_C\ln\!\left[1+{\bf L}_R^\dagger\exp\!\Big(-\frac{E_p+\mu}{T}\Big)\right] \,.
\end{align}
Here, $\mu$ is the chemical potential that is taken to be zero, and $E_p=\sqrt{\vec{p}^2+m_q^2}$ denotes the quark energy where
\begin{align}
\label{eq:constituent-mass}
m_q=g\sigma\,,
\end{align}
is the constituent quark mass. In the PNJL model, the constituent quark masses contain in general,  a linear as well as higher powers of $\sigma$ stemming, for example, from the anomaly fermion operator. In this model, this information is now carried by the scalar potential term. In the fundamental representation with $N=3$ colours, we have
\begin{align}
	S_F &=\ln\!\Big[1+(3\ell-1)e^{-E_p/T}+e^{-2E_p/T}\Big] \notag \\
	&\quad\,+\ln\!\Big(1+e^{-E_p/T}\Big) \,,
	\label{eq:SF}
\end{align}
where, for simplicity, we have implemented the reality condition i.e.~$\ell=\ell^*$. For $N>3$, it requires an additional traced Polyakov loop in $S_F$ expressions.

In a simultaneous expansion in powers of $\ell$ and in the number of meson fields while keeping chiral symmetry manifest, we expect operators of the type  $T^2(g_1\ell+g_2\ell^2)\text{Tr}\left[\Phi^{\dagger }\Phi \right]$ to naturally emerge \cite{Mocsy:2003qw} from the first principle computations. However, given the level of approximations made here, one generates only a mixing between $\ell$ and the scalar field $\sigma$ which might be sufficient for a phenomenological analysis of the thermal phase transitions.  

To summarize, the total effective potential is
\begin{align}
\label{eq:Vtot}
V_{\rm{tot}}=V_{\rm{PLM}}^{\mathrm{(poly)}}+V_{\text{LSM}}+V_{\rm{medium}} \,.
\end{align}
In the next section, we will use the Cornwall-Jackiw-Tomboulis (CJT) method to obtain the finite temperature effective potential of the LSM sector. To avoid double counting the pure gluon sector of the theory is encoded directly in ($V_{\rm{PLM}}$) and the interplay with the scalar mesonic sector by the medium potential. The CJT treatment will be reserved for the purely mesonic sector of the theory.  

\subsection{CJT Method}
\label{sec:CJT}

Cornwall, Jackiw and Tomboulis first proposed a generalized version of effective action $\Gamma\left(\phi,G\right)$ of composite operators \cite{Cornwall:1974vz}, where the effective action not only depends on $\phi(x)$ but also on the propagator $G(x,y)$. This approach is known as the CJT method. In this generalized version, the effective action becomes the generating functional of the two-particle irreducible (2PI) vacuum graphs rather than the conventional 1PI diagrams. The zero temperature version of CJT method has been generalized to finite temperatures by Amelino-Camelia and Pi in \cite{Amelino-Camelia:1992qfe, Amelino-Camelia:1993rvt}. According to \cite{Amelino-Camelia:1993rvt}, the CJT method is equivalent to summing up the infinite class of ``daisy" and ``super daisy" graphs and is thus useful in studying such strongly coupled models beyond the mean-field approximation.

Below, we will use the imaginary time Matsubara formalism. We evaluate the momentum space integrals with the time component $k_4$ replaced by a summation over discrete frequencies because the temporal coordinate is compactified on a circle. Here, $\beta$ is the inverse of the temperature $T$. This leads to the following relationship:
\begin{align}
\int\!\frac{\mathrm d^4k}{\left(2\pi\right)^4}f(k)&\rightarrow \frac{1}{\beta}\sum_n\int\!\frac{\mathrm d^3\vec{k}}{\left(2\pi\right)^3}f\left(2\pi inT,\vec{k}\right)\notag \\
&\equiv\int_\beta f\left(k_4,\vec{k}\right)\,,
\end{align}
where $f$ is a generic function and we have used $k_4=2\pi inT$ in the last step.

According to the CJT formalism \cite{Amelino-Camelia:1993rvt}, the finite temperature effective potential with a generic scalar field $\phi$ is given by:
\begin{align}
&V_{\rm{CJT}}(\phi,G)=V_0(\phi)+\frac{1}{2}\sum_i\int_\beta \ln G_i^{-1}(\phi;k) \notag\\
&+\frac{1}{2}\sum_i\int_\beta\left[D^{-1}(\phi;k)G(\phi;k)-1\right]+V_2(\phi,G)\,,
\label{CJT_formalism}
\end{align}
where the sum runs over all meson species and $V_0$, $D^{-1}\left(\phi;k\right)$, $V_2(\phi,G)$ denote, respectively, the tree level potential, tree level propagator as well as the infinite sum of the two-particle irreducible vacuum graphs. In this work, we use the Hartree approximation according to which  $V_2(\phi,G)$ is simplified to a one ``double bubble" diagram. In the simplest one-meson case, it is proportional to $\left[\int_\beta G\left(\phi;k\right)\right]^2$. We therefore obtain a gap equation by minimizing the above effective potential with respect to the dressed propagator $G_i(\phi;k)$:
\begin{align}
   \frac{1}{2} G_i^{-1}(\phi;k)=\frac{1}{2}D_i^{-1}(\phi;k)+2\frac{\delta V_2(\phi,G)}{\delta G_i(\phi;k) }\,,
   \label{gap}
\end{align}
where $\delta V_2(\phi,G)/\delta G_i(\phi;k)$ denotes the self energy operator. It is convenient to introduce the full field-dependent propagator:
\begin{align}
G_i\left(\phi,k\right)=\frac{1}{k^2+M_i^2}\,,
\label{full_propagator}
\end{align}
where $M\equiv M\left(\phi,k\right)$ is the effective mass which can be interpreted as tree-level mass dressed by the tadpole contributions. By using the full propagator \cref{full_propagator}, the gap equation \cref{gap} yields a set of equations leading to effective temperature-dependent masses (the detailed expressions for general $N_f$ are given in \cref{sec:CJT_general_Nf}). We can easily check that under the Hartree-Fock approximation the effective mass $M$ becomes momentum-independent. It is because in \cref{gap}, the momentum $k$ in the dressed propagator $G$ and tree level propagator $D$ cancel each other while $\frac{\delta V_2(\phi,G)}{\delta G_i(\phi;k) }$ in the Hartree-Fock approximation leads to a term proportional to $\int_\beta G\left(\phi;k\right)$ where the momentum $k$ is integrated out.

Our final results for the CJT improved finite-temperature effective potential of the LSM sector can be written as:
\begin{align}
V_{\rm{eff}}^{\rm{LSM}}\left(\sigma,T\right)=V_0^{\rm{LSM}}\left(\sigma\right)+V_{\rm{FT}}^{\rm{LSM}}\left(\sigma\right)\,,
\end{align}
with
\begin{align}
\label{eq:VFT}
    V_{\rm{FT}}^{\rm{LSM}}\left(\sigma\right)=\frac{T^4}{2\pi^2}\sum_i\left[J_B(R_i^2)-\frac{1}{4}\left(R_i^2-r_i^2\right)I_B(R_i^2)\right]\,,
\end{align}
where
\begin{align}
 \label{eq:thermal-integrals}
J_B(R^2)&=\int_0^\infty \!\mathrm dx\, x^2 \,\ln\left(1-e^{-\sqrt{x^2+R^2}}\right), \notag \\
I_B(R^2)&=2\frac{\mathrm dJ_B(R^2)}{\mathrm dR^2} \notag \\
&=\int_0^\infty \mathrm dx \frac{x^2}{\sqrt{x^2+R^2}}\frac{1}{e^{\sqrt{x^2+R^2}}-1}\,.
\end{align}
Here, $J_B(R^2)$ comes from the logarithmic integral term $\int_\beta \ln G_i^{-1}(\phi,k)$ in \cref{CJT_formalism} while $I_B(R^2)$ comes from the $\int_\beta G\left(\phi;k\right)$ term. Also, we have implemented the definitions $r_i\equiv m_i(\sigma)/T$ and $R_i\equiv M_i(\sigma,T)/T$ where $m_i(\sigma)$ are the tree-level meson masses while $M_i(\sigma,T)$ are the thermal masses (or dressed masses) defined in \cref{full_propagator}. We list the tree-level meson masses below:
\begin{align}
m_{\pi}^2&=-m^2+\frac{1}{2}\lambda_\sigma v^2- \left(c/N_f\right)v^{N_f-2}\,, \notag\\
m_{\eta}^2-m_{\pi}^2&=c\, v^{N_f-2}\,, \notag\\
m_{\sigma}^2-m_\pi^2&=\lambda_\sigma v^2-\left(1-2/N_f\right)c\, v^{N_f-2}\,, \notag\\
m_a^2-m_\pi^2&=\lambda_a v^2+\left(2/N_f\right)c \,v^{N_f-2}\,.
\label{eq:tree_spectrum}
\end{align}
In the chiral limit, i.e., $m_\pi=0$, \cref{eq:tree_spectrum} reduces to
\begin{align}
m_\sigma^2&=\lambda_\sigma v^2-\left(1-2/N_f\right)m_{\eta}^2\,, \notag \\
m_a^2&=\lambda_a v^2 +\left(2/N_f\right) m_\eta^2\,.
\label{eq:spectrum_chiral_limit}
\end{align}

\subsection{Model Parameters and Observables}
\label{sec:pars-obs}

In this section, we summarize all the key model parameters and observables in this work. 

The Polyakov-loop part of the model consists of the Polyakov-loop potential and the interaction term of the Polyakov loop with the LSM. The free parameters in the Polyakov-loop potential \cref{eq:PLM_potential} have been fitted in \cite{Huang:2020crf} to lattice data from \cite{Panero:2009tv} for the pure Yang-Mills theory. The introduction of quarks affects these coefficients. Here, we consider only the impact generated from the medium potential \cref{eq:medium-pot}. We expect the medium potential to capture the main features of the fully non-perturbative results. The medium potential describes the interaction between the Polyakov loop and the LSM and introduces one more free parameter, i.e. the coupling $g$ between the scalar field and the fermions.

For the purely LSM part of the model, for fixed $N_f$, we have four input parameters $\left(m^2,\lambda_\sigma,\lambda_a,c\right)$ which we fix via the four observables $\left(f_\pi,m_\sigma,m_{\eta},m_a\right)$. We are not considering the pion mass since we are working in the chiral limit and therefore the pion mass vanishes, $m_\pi=0$. It should be noted here that the above observables are defined at zero temperature, see \cref{eq:pion_decay_constant,eq:spectrum_chiral_limit}.

\section{Phase Transition and Bubble Nucleation}
\label{sec:phase-transition}

\subsection{Bubble nucleation}
\label{sec:bubble-nucl}

In the case of a first-order phase transition, the latter occurs via the nucleation and expansion of vacuum bubbles. Its dynamics is captured by the bubble nucleation rate. One starts with a computation of the temperature-dependent tunnelling rate between the meta-stable and stable vacuum related to the three-dimensional Euclidean action $S_3(T)$ \cite{Coleman:1977py, Callan:1977pt, Linde:1980tt, Linde:1981zj} via
\begin{align}
	\Gamma(T)=T^4\left(\frac{S_3(T)}{2\pi T}\right)^{\!3/2} e^{-S_3(T)/T} \,.
	\label{eq:decay_rate}
\end{align}
The three-dimensional Euclidean action reads
\begin{align}
	S_3(T)=4\pi\!\int_0^\infty \!\!\mathrm dr\,r^2\!\left[\frac{1}{2}\!\left(\frac{\mathrm d\rho}{\mathrm dr}\right)^{\!2} +V_\text{eff}(\rho,T)\right] \,,
	\label{eq:Euclidean_Action_general}
\end{align}
where $\rho$ denotes a generic scalar field with mass dimension one, $\left[\rho\right]=1$, and $V_\text{eff}$ denotes its effective potential. In our case, the effective potential depends on two scalar fields, the Polyakov loop $\ell$ and the scalar field $\Phi$ (see \cref{meson_field}). Once we choose a vacuum configuration with VEV along the $\sigma$ direction, we can focus on the effective potential only depending on $\ell$ and $\sigma$. In the cases considered here (i.e.~$N_f\geq 3$), we always have a first-order chiral phase transition and the Polyakov loop acts as a spectator. Since we are investigating matter in the fundamental representation of the underling gauge theory there is no proper confinement transition justifying why we do not investigate the latter transition \cite{Mocsy:2003qw}. In this respect, the Polyakov loop works indeed as a spectator field for the chiral transition, even though its properties, such as the spatial two-point function, will be affected by the chiral transition itself \cite{Mocsy:2003qw}. 

The chiral phase transition is described by the scalar field $\sigma$, representing the chiral condensate. When the Polyakov loop is included, we use a mean-field approximation in the Polyakov loop $\ell$. This means that we evaluate the Polyakov loop at the minimum of the effective potential for given values of $\sigma$ and $T$. Thus, the potential becomes a function of only the sigma field, $\left(\sigma,T\right)$, $V_\text{eff}(\sigma, T) = V_\text{eff}(\sigma, T, \ell_\text{min}(\sigma, T))$. This is a good approximation as long as the minimum value of the Polyakov loop does not strongly depend on the sigma field, which is indeed the case in our computations. Based on the above argument, the three-dimensional Euclidean action simplifies to
\begin{align}
	S_3(T)=4\pi\!\int_0^\infty \!\!\mathrm dr\,r^2\!\left[\frac{1}{2}\!\left(\frac{\mathrm d\sigma}{\mathrm dr}\right)^{\!2} +V_\text{eff}(\sigma,T)\right] \,.
	\label{eq:Euclidean_Action_fund}
\end{align} 

The bubble profile is obtained by solving the equation of motion of the action in \cref{eq:Euclidean_Action_fund} and is given by
\begin{align}
	\frac{\mathrm d^2\sigma}{\mathrm dr^2}+\frac{2}{r}\frac{\mathrm d\sigma}{\mathrm dr}=\frac{\partial V_\text{eff}}{\partial\sigma}\,,
	\label{EOM_fund}
\end{align}
with the associated boundary conditions
\begin{align}
	\frac{\mathrm d\sigma(r=0,T)}{\mathrm dr}&=0\,,
	&
	\lim_{r\rightarrow \infty} \sigma(r,T)&=0\,.
 \label{boundary}
\end{align}
To solve \cref{EOM_fund,boundary}, we have implemented the overshooting/undershooting method and employed the \texttt{Python} package \texttt{CosmoTransitions} \cite{Wainwright:2011kj}. We substitute the solved bubble profile $\sigma(r,T)$ into the three-dimensional Euclidean action \cref{eq:Euclidean_Action_fund} and, after integrating over $r$, $S_3$ depends only on $T$.

\subsection{Gravitational-wave parameters}
\label{sec:GW-params}

\subsubsection{Inverse duration time}
\label{sec:beta}

We start with the decay rate of the false vacuum to the true vacuum due to thermal effects (while the decay rate due to quantum corrections is strongly suppressed). For sufficiently fast phase transitions, the decay rate can be approximated by
\begin{align}
	\Gamma(T) \approx \Gamma(t_*) e^{\beta (t-t_*)}\,,
	\label{eq:approx-Gamma}
\end{align}
where $t_*$ is a characteristic time scale for the production of GWs to be specified below. One of the three key parameters in the GW spectrum, the inverse duration time $\beta$ is defined as:
\begin{align}
	\beta= - \frac{\mathrm d}{\mathrm dt} \frac{S_3(T)}{T}\bigg\vert_{t=t_*}\,.
	\label{beta}
\end{align}
It is often convenient to introduce a dimensionless version $\tilde \beta$ which is defined relative to the Hubble parameter $H_*$ at the characteristic time $t_*$ as
\begin{align}
	\tilde \beta = \frac{\beta}{H_*}=T\frac{\mathrm d}{\mathrm dT}\frac{S_3(T)}{T}\bigg\vert_{T=T_*}\,,
	\label{eq:beta-tilde}
\end{align}
where we used that $\mathrm dT/\mathrm dt = -H(T)T$. Note that here, for simplicity, we have assumed that the temperature in the hidden and visible sectors are the same, $T_d=T_v$. But in a more general case, these two temperatures can be different. Introducing a difference between the hidden and visible temperature can in some cases avoid phenomenological constraints \cite{Breitbach:2018ddu, Carenza:2023shd}.

The phase-transition temperature $T_*$ is often identified with the nucleation temperature $T_n$, which is defined as the temperature at which the rate of bubble nucleation per Hubble volume and time is approximately one, i.e.\ $\Gamma/H^4\sim \mathcal{O}(1)$. More accurately, one can use the percolation temperature $T_p$, which is defined as the temperature at which the probability of being in the false vacuum is about $0.7$. It is apparent that the percolation temperature corresponds to a point when the phase transition has proceeded further compared with that at the nucleation temperature, and thus $T_p\leq T_n$. We use the percolation temperature throughout this work. To find it, we first write the false-vacuum probability as~\cite{Guth:1979bh, Guth:1981uk}
\begin{align}
	P(T) = e^{-I(T)}\,,
\end{align}
with the weight function \cite{Ellis:2018mja}
\begin{align}
	I(T)=\frac{4\pi}{3} \int^{T_c}_T \! \!\mathrm dT'\frac{\Gamma(T')}{H(T')T'{}^4} \left( \int^{T'}_{T}\!\!\mathrm dT''\frac{v_w(T'')}{H(T'')}\right)^{\!3} \,.
	\label{eq:Tp}
\end{align}
The percolation temperature is defined by $I(T_p)=0.34$, corresponding to $P(T_p)= 0.7$ \cite{Rintoul_1997}. Using $T_*=T_p$ in \cref{eq:beta-tilde} yields the dimensionless inverse duration time.

\subsubsection{Energy budget}
\label{sec:budget}

For the second important parameter the phase transition strength $\alpha$, there are two different definitions. In the literature, $\alpha$ is often defined as the latent heat during the phase transition per d.o.f. In this work, we define the strength parameter $\alpha$ by using the trace of the energy-momentum tensor $\theta$ weighted by the enthalpy
\begin{align}
	\alpha=\frac{1}{3}\frac{\Delta\theta}{w_+}=\frac{1}{3}\frac{\Delta e\,-\,3\Delta p}{w_+}\,,
	\label{alpha_def}
\end{align}
where $\Delta X= X^{(+)}-X^{(-)}$ for $X = (\theta$, $e$, $p$) and $(+)$ denotes the meta-stable phase (outside of the bubble) while $(-)$ denotes the stable phase (inside of the bubble). This definition quantifies the jump of the energy density and the pressure across the phase boundary weighted by the enthalpy and thus it measures the strength of the first-order phase transition. The enthalpy quantifies the total d.o.f.'s of the system which participates in the phase transition and the relations between enthalpy $w$, pressure $p$, and energy $e$ are given by
\begin{align}
	w&=\frac{\partial p}{\partial \ln T}\,,
	&
	e&=\frac{\partial p}{\partial \ln T} -p\,.
	\label{eq:enthalpy}
\end{align}
These are hydrodynamic quantities and we work in the approximation where do not solve the hydrodynamic equations but instead extract them from the effective potential with
\begin{align}
	p^{(\pm)}&=-V_{\text{eff}}^{(\pm)}\,.
	\label{eq:def-p}
\end{align}
This treatment should work well for the phase transitions considered here, see \cite{Giese:2020rtr, Giese:2020znk, Wang:2020nzm}. With \cref{eq:enthalpy,eq:def-p}, $\alpha$ is given by
\begin{align}
	\alpha=\frac{1}{3}\frac{4\Delta V_\text{eff}-T\frac{\partial \Delta V_\text{eff}}{\partial T}}{-T\frac{\partial V_{\text{eff}}^{(+)}}{\partial T}}\,.
	\label{eq:alpha}
\end{align}
Note that relativistic SM d.o.f.'s do not contribute to our definition of $\alpha$ since they are fully decoupled from the phase transition. The dilution due to the SM d.o.f.'s is included at a later stage, see \cref{sec:spectrum}.

\subsubsection{Bubble-wall velocity}
\label{sec:wall-velocity}

We treat the bubble-wall velocity $v_w$ as a free parameter. A reliable estimate of the wall velocity would require a detailed analysis of the pressure and friction on the bubble wall. The latter is typically evaluated in an expansion of $1 \to n$ processes \cite{Bodeker:2009qy, Bodeker:2017cim, Cai:2020djd, Baldes:2020kam, Azatov:2020ufh, Wang:2020zlf}. In our case, we have a strongly-coupled system and most likely a fully non-perturbative analysis would be necessary to determine the friction. Some initial works towards this direction by using holographic methods can be found in \cite{Bigazzi:2021ucw, Bea:2021zsu, Janik:2022wsx, Chen:2022cgj}.

We display results for two different wall velocities, once the Chapman-Jouguet (CJ) detonation
velocity, which is given by
\begin{align}
 v_J=\frac{\sqrt{2\alpha/3+\alpha^2}+\sqrt{1/3}}{1+\alpha} \,,
\end{align}
and once for $v_w=0.1$. Using the CJ velocity is an optimistic assumption since it leads to the largest efficiency factor for the production of GWs, see next section, and therefore also to the largest GW signal. The results with the CJ velocity can be interpreted as an upper bound for the true GW signal. Note however, that as long as the wall velocity is larger than the speed of sound, $v_w \geq c_s=1/\sqrt{3}$, the wall velocity does not strongly impact the GW peak amplitude. For wall velocities smaller than the speed of sound, the efficiency factor decreases rapidly and the generation of GW from sound waves is suppressed~\cite{Cutting:2019zws}.

\subsubsection{Efficiency factors}
\label{sec:eff-factors}

The efficiency factors determine which fraction of the energy budget is converted into GWs. In this work, we focus on the GWs from sound waves, which is the dominating contribution to the phase transitions considered here. The efficiency factor for the sound waves $\kappa_\text{sw}$ consists of the factor $\kappa_v$ \cite{Espinosa:2010hh} as well as an additional suppression due to the length of the sound-wave period $\tau_\text{sw}$ \cite{Ellis:2019oqb, Ellis:2020awk, Guo:2020grp}
\begin{align}
	\kappa_\text{sw}&= \sqrt{\tau_\text{sw}} \, \kappa_v\,.\label{eq:efficiency}
\end{align}
In our notation, $\tau_\text{sw}$ is dimensionless and measured in units of the Hubble time. It is given by \cite{Guo:2020grp}
\begin{align}
	\tau_\text{sw}=1-1/\sqrt{1+2\frac{(8\pi)^{\frac 13} v_w}{\tilde \beta \,\bar U_f}}\,
 \label{eq:suppression}
\end{align}
and for $\tilde\beta\gg 1$, $\tau_\text{sw}$ can be simplified to
\begin{equation}
   \tau_\text{sw}\sim\frac{(8\pi)^{\frac 13} v_w}{\tilde \beta \,\bar U_f}\,.
    \label{eq:suppression_simplified}
\end{equation}
It is apparent that for larger $\beta$, $\tau_\text{sw}$ will be suppressed. $\bar U_f$ is the root-mean-square fluid velocity \cite{Hindmarsh:2015qta, Ellis:2019oqb}
\begin{align}
	\bar U_f^2 = \frac{3}{v_w(1+\alpha)}\int^{v_w}_{c_s}\!\mathrm d\xi \, \xi^2 \frac{v(\xi)^2}{1-v(\xi)^2}\simeq \frac{3}{4}\frac{\alpha}{1+\alpha}\kappa_v \,.
\end{align}
We follow \cite{Espinosa:2010hh} for $\kappa_v$ where it was numerically fitted to simulation results. The factor $\kappa_v$ depends on $\alpha$ and $v_w$, and, for example, at the CJ velocity it reads
\begin{align}
	\kappa_v(v_w=v_J)=\frac{\sqrt{\alpha}}{0.135 +\sqrt{0.98+\alpha}}\,.
	\label{eq:eff-vJ}
\end{align}
For a phase transition with $\alpha \sim \mathcal{O}(10^{-2})$, we have $\kappa_v \sim 0.1$.

\subsection{Gravitational-wave spectrum}
\label{sec:spectrum}

To extract the GW spectrum from the parameters $\alpha$, $\tilde \beta$, we follow the standard procedures in \cite{Caprini:2015zlo, Caprini:2019egz}. The contributions to the GW signals from bubble collision \cite{Kosowsky:1991ua, Kosowsky:1992rz, Kosowsky:1992vn, Kamionkowski:1993fg, Caprini:2007xq, Huber:2008hg, Caprini:2009fx, Espinosa:2010hh, Weir:2016tov, Jinno:2016vai} and magnetohydrodynamic turbulence in the plasma \cite{Kosowsky:2001xp, Dolgov:2002ra, Caprini:2006jb, Gogoberidze:2007an, Kahniashvili:2008pe, Kahniashvili:2009mf, Caprini:2009yp, Kisslinger:2015hua} are subleading compared to the sound waves in the considered regime \cite{Hindmarsh:2013xza, Giblin:2013kea, Giblin:2014qia, Hindmarsh:2015qta, Hindmarsh:2017gnf}. Thus, in the following, we focus on the contributions from sound waves in the plasma only.

\begin{figure*}[tbp]
    \includegraphics[width=\linewidth]{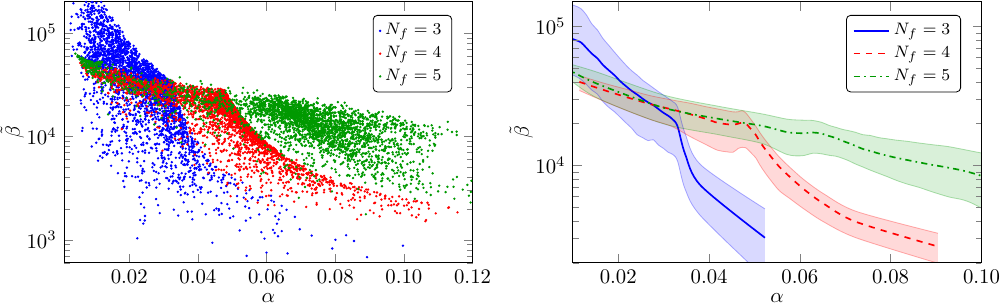}
    \caption{We show the range of $\alpha$ and $\tilde \beta$ values of the LSM for $N_f=3,4,5$. In the left panel, we show the actual distribution of theory points, while in the right panel, we display the averaged values. On average, the LSM produces stronger GW signals with increasing $N_f$ due to the larger $\alpha$ values. Nonetheless, the strongest GW signals are achieved with the LSM for $N_f=3$, corresponding to the sparse blue dots at small $\tilde\beta$ in the left panel.}
     \label{fig:LSM_alpha-beta}
\end{figure*}

The GW spectrum sourced by sound waves in the plasma is given by
\begin{align}
	h^2\Omega_\text{GW}(f)&= h^2\Omega^\text{peak}_\text{GW} \left(\frac{f}{f_\text{peak}}\right)^{\!3} \left[ \frac{4}{7}+\frac{3}{7}\left( \frac{f}{f_\text{peak}} \right)^{\!2}\right]^{-\frac{7}{2}}\!,
	\label{eq:GWsignal}
\end{align}
with the peak frequency
\begin{align}
	f_\text{peak}&\simeq 1.9\cdot 10^{-5}\,\text{Hz}\left(\frac{g_*}{100} \right)^{\!\frac{1}{6}}\left( \frac{T}{100\, \text{GeV}}\right) \left(\frac{\tilde \beta}{v_w} \right) \,,
	\label{eq:peak-f}
\end{align}
and the peak amplitude
\begin{align}\label{eq:peak-amp}
	h^2\Omega^\text{peak}_\text{GW} &\simeq 2.65\cdot 10^{-6}\left(\frac{v_w}{\tilde \beta}\right)\left( \frac{\kappa\, \alpha}{1+\alpha} \right)^{\!2}\left(\frac{100}{g_*}\right)^{\!\frac{1}{3}}\Omega_\text{dark}^2\,.
\end{align}
Here, $h= H/(100 \text{km}/\text{s}/\text{Mpc})$ is the dimensionless Hubble parameter and $g_*$ is the effective number of relativistic d.o.f.'s, including the SM d.o.f.'s \ $g_{*,\text{SM}}=106.75$ and the dark sector ones, which is $g_{*,\text{dark}} = 16+2N_f^2$, where the 16 comes from the $SU(3)$ gluon d.o.f.'s~while $2N_f^2$ comes from the meson field $\Phi$ (see \cref{meson_field} where $N^2-1$ d.o.f.'s,~respectively, for $a$ field and $\pi$ field and 2 extra d.o.f.'s~for $\sigma$ and $\eta'$).

The factor $\Omega_\text{dark}^2$ in \cref{eq:peak-amp} accounts for the dilution of the GWs by the visible SM matter which does not participate in the phase transition. The factor reads
\begin{align}
	\label{eq:dilution}
	\Omega_\text{dark} =\frac{\rho_{\text{rad},\text{dark}}}{\rho_\text{rad,tot}}=\frac{g_{*,\text{dark}} }{g_{*,\text{dark}}+g_{*,\text{SM}}}\,.
\end{align}
It is also possible to absorb the above dilution factor into the redefinition of the strength parameter (see e.g.~\cite{Breitbach:2018ddu,Huang:2020crf}). The decisive quantity that determines the detectability of a GW signal is the signal-to-noise ratio (SNR) \cite{Allen:1997ad, Maggiore:1999vm} given by
\begin{align}
	\text{SNR} = \sqrt{\frac{T}{\text{s}} \int_{f_\text{min}}^{f_\text{max}}\mathrm df \left(\frac{ h^2 \Omega_\text{GW}}{h^2 \Omega_\text{det}} \right)^{\!2}}\,.
	\label{eq:SNR}
\end{align}
Here, $h^2 \Omega_\text{GW}$ is the GW spectrum given by \cref{eq:GWsignal}, $h^2 \Omega_\text{det}$ is the sensitivity curve of the detector, and $T$ is the observation time, for which we assume $T=3$\,years. We compute the SNR of the GW signals for the future GW observatories LISA \cite{Audley:2017drz, Baker:2019nia, LISAdocument}, BBO \cite{Crowder:2005nr, Corbin:2005ny, Harry:2006fi, Thrane:2013oya, Yagi:2011wg}, and DECIGO \cite{Seto:2001qf, Yagi:2011wg, Kawamura:2006up, Isoyama:2018rjb}. The sensitivity curves of these detectors are nicely summarised and provided in \cite{Schmitz:2020syl}.
\begin{figure*}[tbp]
    \includegraphics[width=\linewidth]{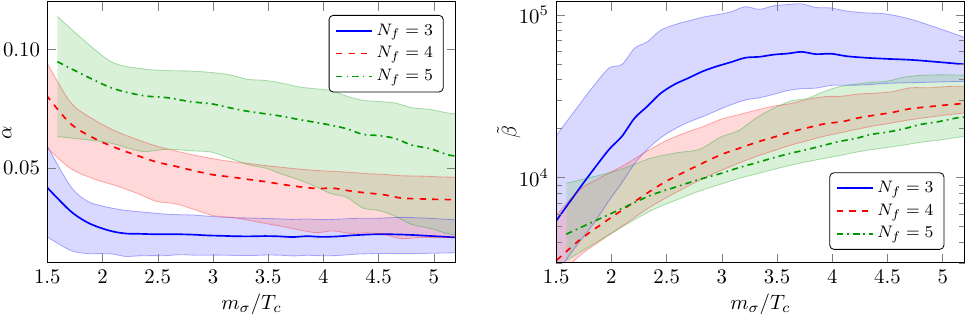}
    \caption{We show the averaged values of $\alpha$ (left panel) and $\tilde\beta$ (right panel) in the LSM for $N_f=3,4,5$ as a function of the sigma meson mass in units of $T_c$. Smaller values of $m_\sigma$ correspond to larger values of $\alpha$ and smaller values of $\tilde \beta$ leading to a stronger first-order phase transition.}
    \label{fig:LSM_msigma-alpha-beta}
\end{figure*}

\section{Results}
\label{sec:results}

We are now ready to present the results starting with the LSM as a benchmark model for $N_f= 3,4$, and 5. For $N_f=3$, we compare the PLSM and the LSM while for $N_f=5$ we use LSM alone but consider separately the cases in which the determinant operator is turned on or off in \cref{VLSM}.

For all these models, we randomly choose values for the meson masses and the pion decay constant. Afterwards, the ratios are adjusted so the phase transition temperature is at $T_c = 1$\,GeV. The values of the mass parameters are chosen from the range
\begin{align}
    \label{eq:obs-choice}
    f_\pi &\in (20,\, 200)\,\text{MeV} \,, \notag \\
    m_\sigma &\in (0.1,\, 1)\,\text{GeV} \,, \notag  \\
    m_\eta &\in (0.1,\, 1)\,\text{GeV} \,, \notag  \\
    m_a &\in (0.1,\, 1)\,\text{GeV} \,.
\end{align}
Note that we are working in the chiral limit and thus $m_\pi=0$. After these values are chosen, we relate them to the Lagrangian parameters of the theory $(m^2,\,\lambda_\sigma,\,\lambda_a,\,c)$ using \cref{eq:spectrum_chiral_limit}. We then perform some sanity checks on these theories: firstly, we demand that the vacuum is stable. The vacuum stability condition reads \cite{Hansen:2017pwe}
\begin{align}
    \label{eq:vac-stability}
 \lambda_a > - \frac{1}{N_f-1} \lambda_\sigma\,. 
\end{align}
Secondly, we demand that the theory has a broken chiral symmetry at zero temperature. Lastly, we limit the magnitude of the couplings to be, in absolute value, within the range $\lambda_a,\lambda_\sigma < 16\pi$. If these demands are not met we discard the given theory point. 

Around 60k values are randomly chosen for each model in the ranges \cref{eq:obs-choice}, and all relevant phase transition parameters are computed. In this manner, we obtain plots such as the left panel in \cref{fig:LSM_alpha-beta}, where the phase transition parameters $\alpha$ and $\tilde\beta$ are displayed for the LSM with $N_f=3,4,5$. Each point in the left panel of \cref{fig:LSM_alpha-beta} represents a theory obtained for a given choice of input parameters. Due to the large amount of randomly chosen parameter values, we obtain a thorough picture of the entire $(\alpha,\tilde\beta)$  LSM landscape. Note that for readability reasons, we have reduced the number of points shown to three thousand per given number of flavours in the left panel of \cref{fig:LSM_alpha-beta}.

\begin{figure*}[tbp]
    \includegraphics[width=\linewidth]{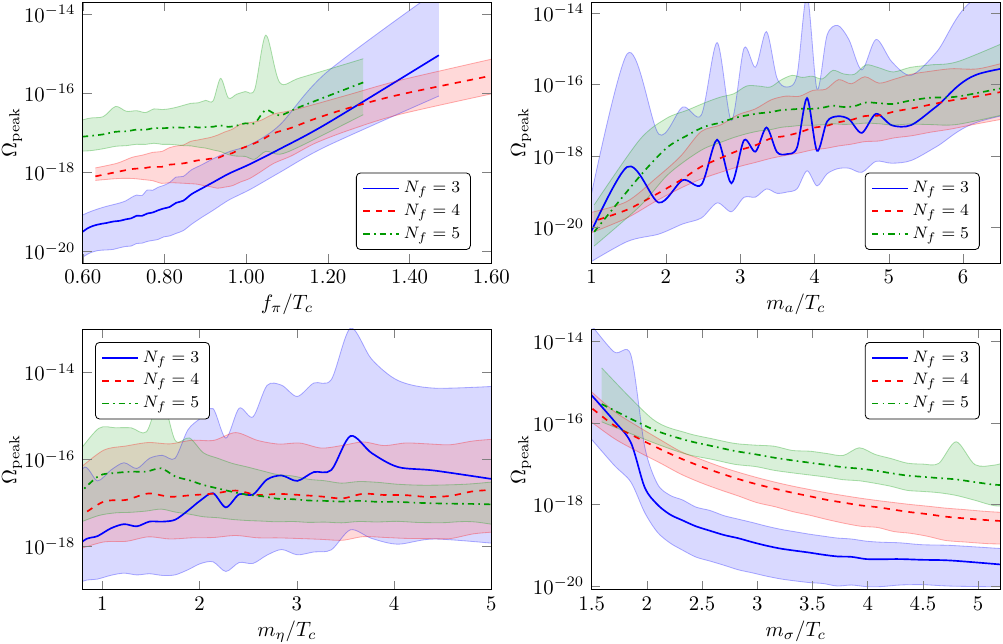}
    \caption{We show the averaged values of the peak amplitude $\Omega_\text{peak}$ as a function of physical observables $f_\pi$ (top left), $m_\sigma$ (top right), $m_\eta$ (bottom left), and $m_a$ (bottom right) all in units of $T_c$ in the LSM for $N_f=3,\, 4,\,5$. The pion decay constant and the sigma meson mass have the strongest correlation with the peak amplitude: larger values of $f_\pi$ and smaller values of $m_\sigma$ lead to a larger $\Omega_\text{peak}$.}
    \label{fig:LSM_peak-obs_binned}
\end{figure*}

While the plots give us a thorough picture of what a given underlying theory can achieve, they are not immediate to interpret. Therefore, we create more accessible plots by averaging over the points. For example, we sort the results in $\alpha$ and then group them in sets of roughly 2k points per set. We compute the mean and assume for the variance a Gau\ss ian distribution above and below the mean. This leads to the plot in the right panel of \cref{fig:LSM_alpha-beta}. While this process washes out some features of the full data it still displays the important qualitative features. For example, the $\alpha-\tilde\beta$-data displays a characteristic 'edge' around $(\alpha,\tilde\beta)= (0.033, 3\cdot 10^4)$ for $N_f=3$ and $(\alpha,\tilde\beta)= (0.048, 3\cdot 10^4)$ for $N_f=4$. This edge is also visible in the averaged data in \cref{fig:LSM_alpha-beta}. We made sure that the averaging process did not wash out important physics features in the results that we displayed.

\subsection{Flavour dependence}
\label{sec:flavour-dependence}

In \cref{fig:LSM_alpha-beta}, we show the averaged plots of the $\tilde \beta$ vs $\alpha$ plane for $N_f=3,4,5$. The strongest first-order phase transitions are in the bottom right corner of these plots, where $\alpha$ is large and $\tilde \beta$ is small. We find that on average the first-order phase transitions are stronger for larger $N_f$ due to the larger $\alpha$ values (the curves are shifted towards the right with increasing $N_f$). Note however, that the strongest first-order phase transitions occur in the $N_f=3$ case: there are a few sparse blue points in the left panel of \cref{fig:LSM_alpha-beta} with $\tilde \beta \sim 10^3$. These happen due to a zero-temperature barrier in the potential which is induced from the determinantal term going $\sim \sigma^3$ for $N_f=3$. Note that this cannot happen for other values of $N_f$. To support the statement of \cref{fig:LSM_alpha-beta}, we show in \cref{fig:LSM_msigma-alpha-beta} the GW parameters $\alpha$ and $\tilde\beta$ separately as a function of the sigma meson mass $m_\sigma$. The sigma meson mass is the best parameter to plot against as we will show below. In \cref{fig:LSM_msigma-alpha-beta}, we can clearly see that $\alpha$ gets larger (left panel) and $\tilde \beta$ gets smaller (right panel) with increasing $N_f$, both leading to a stronger GW signal.

\begin{figure*}[tbp]
    \includegraphics[width=\linewidth]{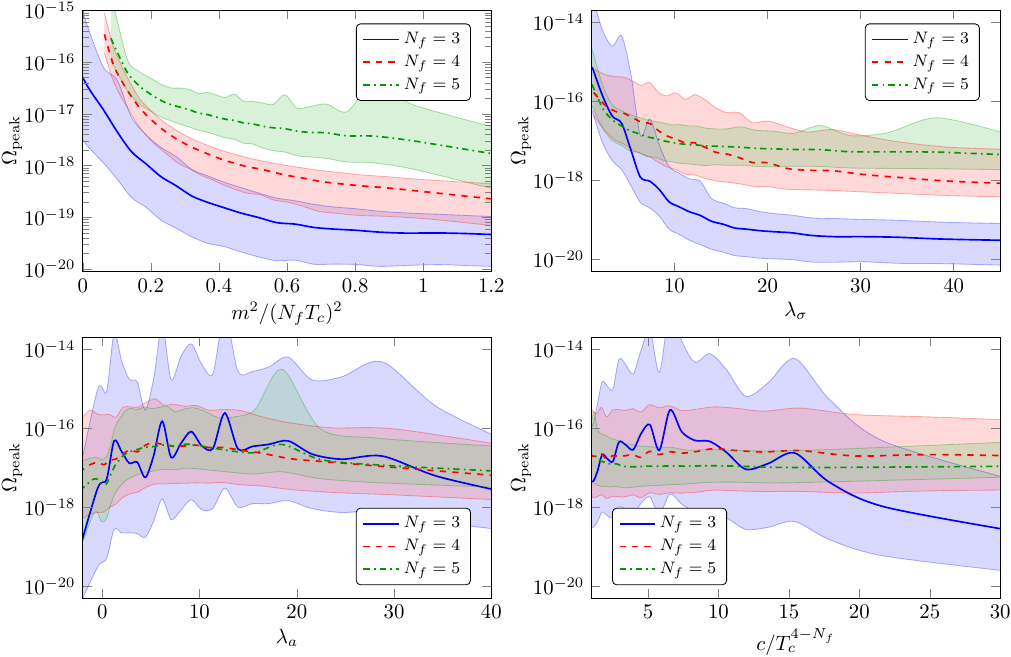}
    \caption{We show the averaged values of the peak amplitude $\Omega_\text{peak}$ as a function of the model parameters $m^2$ (top left), $\lambda_\sigma$ (top right), $\lambda_a$ (bottom left), and $c$ (bottom right) in the LSM for $N_f=3,\, 4,\,5$. The strongest correlation is found between $\Omega_\text{peak}$ and $m^2$ as well as $\lambda_\sigma$: smaller values of $m^2$ and $\lambda_\sigma$ lead to a larger $\Omega_\text{peak}$.}
    \label{fig:LSM_peak-coupling_binned}
\end{figure*}

\subsection{Dependence on physical and theory parameters}
\label{sec:dependence-parameters}

We now want to understand what are the dependencies on the physical parameters (masses and pion decay constant) as well as on the theoretical model parameters. For this, we compute the peak amplitude, \cref{eq:peak-amp}, with the assumption that wall velocity is given by the Chapman-Jouguet detonation velocity \cref{eq:eff-vJ}. This is an optimistic assumption as it maximises the peak amplitude and most likely the strong gluon dynamics lead to significant friction and a slower wall velocity \cite{Asadi:2021yml, Asadi:2021pwo, Bigazzi:2021ucw, Bea:2021zsu, Janik:2022wsx}. Nonetheless, we use this as a benchmark and discuss the impact of the wall velocity later in \cref{sec:SNR}. Overall, the peak amplitude gives us the most direct relation to the strength of the first-order phase transition, and therefore we plot it as a function of the physical and theory parameters.

In \cref{fig:LSM_peak-obs_binned}, we show the averaged plots of the peak amplitude $\Omega_\text{peak}$ as a function of the pion decay constant $f_\pi$ and the masses $m_\sigma,\,m_\eta,\,m_a$. The latter are all measured in units of $T_c$. We observe that the strongest correlation are between $\Omega_\text{peak}$ and $f_\pi$ as well as $m_\sigma$: $\Omega_\text{peak}$ increases with increasing $f_\pi$ and with decreasing $m_\sigma$. In particular, the width of the curves is the smallest for $m_\sigma$ indicating that the distribution of theories is most clearly sorted with this parameter. Note, that for $N_f=3$ and small $m_\sigma$, the largest GW amplitudes are generated, reaching up to $\Omega_\text{peak} = \mathcal{O}(10^{-14})$, which would be almost in reach of the sensitivity of LISA. Due to the clear correlation between $\Omega_\text{peak}$ and $m_\sigma$, we will use $m_\sigma$ as the preferred plotting parameter in the upcoming sections.

On the other hand, we only see a mild dependence of the peak amplitude on the masses $m_\eta$ and $m_a$. In particular, there is almost no dependence on $m_\eta$, while we observe a mild increase of $\Omega_\text{peak}$ with $m_a$. The $N_f=3$ curves are oscillating strongly as a function of $m_\eta$ and especially as a function of $m_a$. The reason is simply that the curves have not converged yet despite us having included 60k theory points. This is due to the very wide distributions as a function of these parameters.

In \cref{fig:LSM_peak-coupling_binned}, we show the averaged plots of the peak amplitude $\Omega_\text{peak}$ as a function of the couplings $m^2$, $\lambda_\sigma$, $\lambda_a$, and $c$. The parameter $m^2$ is measured in units of $N_f T_c$ and $c$ is measured in units of $T_c$ if it is dimensionful. 

We observe a very similar picture as with the physical parameters: we have two parameters that display a strong correlation ($m^2$ and $\lambda_\sigma$) and two parameters that display almost no correlation ($\lambda_a$ and $c$). The peak amplitude increases with a decreasing $m^2$ and a decreasing $\lambda_\sigma$. The strongest is between $\Omega_\text{peak}$ and $\lambda_\sigma$, which is in straight analogy to the correlation with $m_\sigma$. This can be understood directly at the hand of the relation between $\lambda_\sigma$ and $m_\sigma$, as we will discuss in \cref{Sec:discussion}.

Overall, our observations are consistent with the results discussed in \cref{sec:flavour-dependence} since the GW peak amplitude generically increases with $N_f$, which is most easily visible in the top two panels of both, \cref{fig:LSM_peak-obs_binned,fig:LSM_peak-coupling_binned}.

\begin{figure*}[tbp]
    \includegraphics[width=\linewidth]{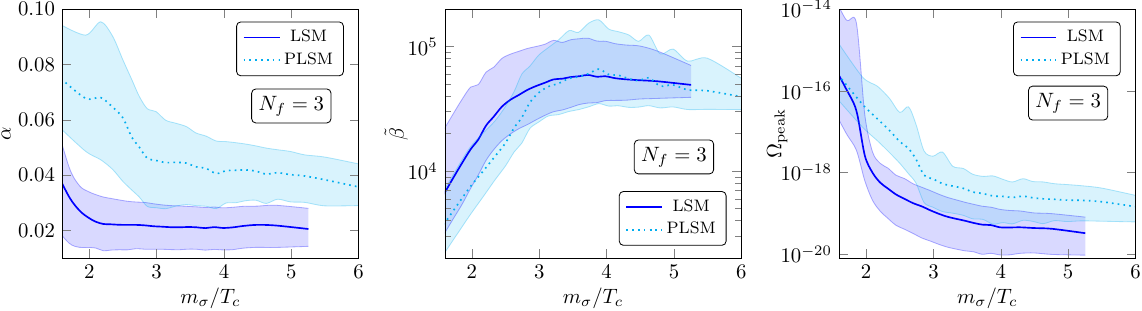}
    \caption{We show the impact of the Polyakov loop on the GW parameters $\alpha$ (left panel), $\tilde\beta$ (central panel), and on the peak amplitude $\Omega_\text{peak}$ as a function of the sigma meson mass $m_\sigma$ at the example of $N_f=3$ and the coupling $g=1$ in the constituent quark mass, \cref{eq:constituent-mass}. While the Polyakov loop has no effect on $\tilde \beta$, it increases the strength of the phase transition by increasing the value of $\alpha$.}
    \label{fig:PLSM}
\end{figure*}

\begin{figure*}[tbp]
    \includegraphics[width=\linewidth]{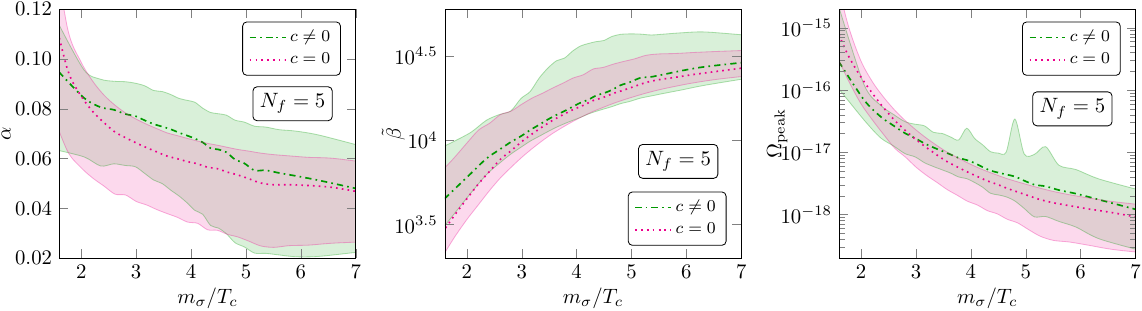}
    \caption{We show the impact of the determinant term on the GW parameters $\alpha$ (left panel), $\tilde\beta$ (central panel), and on the peak amplitude $\Omega_\text{peak}$ as a function of the sigma meson mass $m_\sigma$ at the example of the LSM for $N_f=5$. Although the determinant term is non-renormalisable for $N_f=5$, there is no effect on the strength of the GW signal.}
    \label{fig:determinant}
\end{figure*}

\subsection{PLSM vs LSM}
\label{sec:res-PLSM}

In this section, we investigate the effect of the Polyakov loop on the GW spectrum. This implies that we include the Polyakov loop potential \cref{eq:PLM_potential} and the medium potential \cref{eq:medium-pot} to the LSM. This results in one new free parameter, the coupling $g$ in the constituent quark mass \cref{eq:constituent-mass}, which we set to $g=1$ for this investigation. We also restrict ourselves to $N_f=3$ but we expect that the inclusion of the Polyakov loop has similar overall effects for all $N_f$. We note that the inclusion of the Polyakov loop increases the complexity of the computation and therefore we use fewer statistics for the PLSM (14k points versus 60k points), which results in a less converges and less monotonic plot for the PLSM.

In \cref{fig:PLSM}, we show the GW parameters $\alpha$ and $\tilde \beta$ as well as the peak amplitude $\Omega_\text{peak}$ as a function of the sigma-meson mass $m_\sigma$ in the PLSM and LSM for $N_f=3$. We observe that the Polyakov loop has little effect on the parameter $\tilde \beta$ (central panel) while it increases the strength parameter $\alpha$ compared to the LSM (left panel). In summary, this leads to an increased GW peak amplitude in the PLSM (right panel). Note, however, that the strongest GW signals are still present for the LSM at very small values of the sigma meson mass. 

\begin{figure*}[tbp]
    \includegraphics[width=\linewidth]{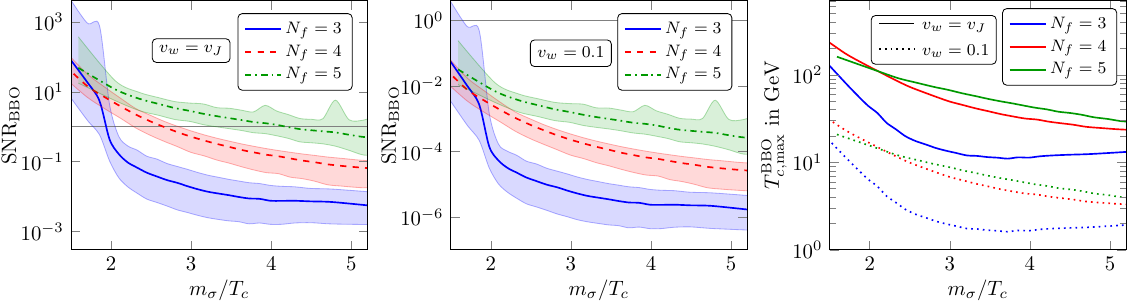}
    \caption{Signal-to-noise ratio at BBO as a function of the sigma-meson mass for the optimistic case that the wall velocity is $v_w=v_J(\alpha)$ (left panel) and for the wall velocity $v_w=0.1$ (central panel). We choose the critical temperature of the phase transition such that the peak frequency falls on the maximal sensitivity of the BBO detector. The critical temperature that maximises the signal-to-noise ratio is shown in the right panel.}
    \label{fig:SNR}
\end{figure*}

\subsection{Determinant operator}
\label{sec:determinant}

In this section, we investigate the influence of the determinant operator. This is in particular relevant for larger $N_f$ since the mass dimension of the determinant term is $[c] = 4 - N_f$, which means that it is a relevant operator for $N_f=3$, marginal for $N_f =4$, and non-renormalisable for $N_f=5$. Neglecting the determinant term leads to a vanishing mass for the eta-meson, $m_\eta=0$. To ensure a fair comparison between the theories, we included the determinant term for $N_f=5$ in the previous sections. Since for $N_f=5$, the term is non-renormalisable this leads to an un- or meta-stable tree-level potential and we interpret the theory to be valid below a given cutoff below. The instabilities of the potential are above this given cutoff scale. We discard theories that have a cutoff scale which is too close to the critical temperature. To ensure that this is a valid treatment, we compare in this section the results to the case of a vanishing determinant term.

In \cref{fig:determinant}, we compare the LSM at $N_f=5$ with and without the determinant term. We show the GW parameters $\alpha$ (left panel) and $\tilde \beta$ (central panel) as well as the peak amplitude $\Omega_\text{peak}$ (right panel) as a function of the sigma-meson mass $m_\sigma$. We observe that on average curves agree very well with each other, and we conclude that the determinant term does not influence the strength of the GW signal. This justifies our treatment in the previous sections and also agrees with our observations in \cref{fig:LSM_peak-obs_binned,fig:LSM_peak-coupling_binned} where we have observed that the GW peak amplitude neither depends on $m_\eta$ nor on $c$. This also holds true for $N_f=3$ and 4. The latter is indeed surprising since the determinant term is a relevant operator for $N_f=3$ and it is also responsible for the strongest first-order phase transitions due to a zero temperature barrier in the potential, as shown in \cref{fig:LSM_peak-obs_binned} for small $m_\sigma$ and as discussed in \cref{sec:flavour-dependence}.

\subsection{Signal-to-noise ratio}
\label{sec:SNR}

The signal-to-noise ratio \cref{eq:SNR} is the key quantity to decide whether a GW signal is detectable or not. We focus on BBO since it has the greatest sensitivity of the detectors and assume that the signal is detectable for SNR\,$>1$. We further choose the critical temperature of the phase transition such that the peak frequency falls on the maximal sensitivity of the detector. This implies for BBO a critical temperature of the order of $T_c = \mathcal{O}(100)$\,GeV for $v_w=v_J$ and  $T_c = \mathcal{O}(10)$\,GeV for $v_w=0.1$, which we show in the right panel of \cref{fig:SNR}.

Here we already indicated that the result depends on the terminal wall velocity, which we use as an external parameter since it is difficult to compute in a non-perturbative setting. There have been some indications that the wall velocity is rather small in these kinds of phase transitions since the strong gluon dynamics could provide a lot of friction that slows down the bubble wall. Here, we exemplify our results for two different wall velocities: the Chapman-Jouguet detonation velocity\cref{eq:eff-vJ}, which leads to the strongest GW signal and is an optimistic best-case scenario, and secondly $v_w=0.1$. 

The results are displayed in \cref{fig:SNR}. In the left panel, we show the SNR for $N_f=3,4,5$ as a function of the sigma-meson mass in the best-case scenario with $v_w=v_J$. Here, we find a detectable GW signal for $m_\sigma/T_c <  1.96$ ($N_f=3$), $m_\sigma/T_c <  2.65$ ($N_f=4$), and $m_\sigma/T_c <  4.19$ ($N_f=5$). For smaller wall velocities, $v_w=0.1$, see central panel, none of the GW signals are detectable, with the exception of the very strong first-order phase transitions in the $N_f=3$ case for small $m_\sigma$. 

\section{Discussions: conditions for a stronger first-order phase transition}
\label{Sec:discussion}

In the previous section, we have elucidated the phase structure of the (P)LSM and have identified corners of the theories where strong first-order phase transitions take place, some measurable at future GW detectors. We found the strongest first-order phase transitions for small sigma masses $m_\sigma$, large pion decay constants $f_\pi$, small $m^2$, and small $\lambda_\sigma$. Now we want to discuss how these conditions are interrelated.

\begin{figure*}[tbp]
    \includegraphics[width=\linewidth]{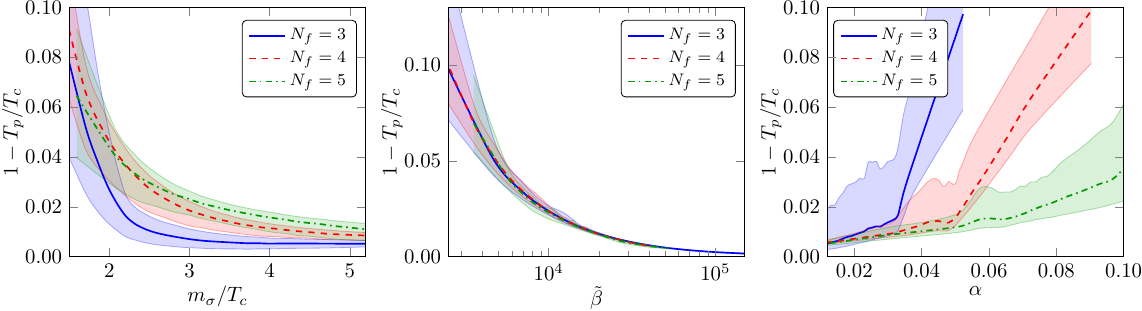}\\
    \caption{Supercooling parameter $1-T_p/T_c$ as a function of the sigma mass $m_\sigma$ in units of $T_c$ (left panel), and the GW parameters $\tilde \beta$ (central panel) and $\alpha$ (right panel).}
    \label{fig:LSM_peak-cooling_binned}
\end{figure*}

Naively, a smaller $\lambda_\sigma$ leads to a larger $m_\sigma$, which is contradictory to our above results. This is most easily seen at the hand of the tree-level expression, see \cref{eq:tree_spectrum}, and utilising  the minimum condition of the tree-level effective potential. For $N_f=3$, we find
\begin{align}
 m_\sigma^2 &= \frac{c \sqrt{c^2+18 \lambda_\sigma m^2}+c^2+18 \lambda_\sigma m^2}{9 \lambda_\sigma} \,,
 \label{eq:msigma-tree}
\end{align}
which shows that a smaller $\lambda_\sigma$ leads to a larger $m_\sigma$. Similar expressions hold for larger $N_f$.

Opposing this naive view, our full data shows a different correlation: we chose the physical parameters randomly in the ranges given in \cref{eq:obs-choice} from which we can directly compute the couplings of the theory. In this data set, we observe the correlation that larger $\lambda_\sigma$ leads to a larger $m_\sigma$.

The reason that the naive expectation from equation \cref{eq:msigma-tree} fails, is that it is based on an analysis for fixed values of $m^2$ and $c$. For a decreasing $\lambda_\sigma$ at fixed $m^2$ and $c$, the other masses ($m_a$ and $m_\eta$) would increase as well and leave the range given in \cref{eq:obs-choice}.

It is more useful to express \cref{eq:msigma-tree} in terms of other physical observables. For example, the equation can be recast in the form ($N_f=3$)
\begin{align}
 m_\sigma^2 &= m_\eta^2/3+ 2m^2\,.
\end{align}
from which it becomes apparent that small $m^2$ corresponds to small $m_\sigma^2$ for fixed $m_\eta^2$. The latter is a valid assumption since $m_\eta$ is dominantly determined by $c$, see \cref{eq:obs-choice}. Similarly, the relation ($N_f=3$)
\begin{align}
    f_\pi & = \sqrt{\frac{2}{3}} \frac1c \, m_\eta^2\,,
\end{align}
can be misleading, since it does not imply a larger $f_\pi$ for smaller $c$. Instead, a smaller $c$ corresponds to a smaller $m_\eta^2$, which together keeps $f_\pi$ constant.

More straightforwardly, a large  $f_\pi$ corresponds to a large vev, since
\begin{align}
    f_\pi & = \sqrt{\frac2{N_f}}\,v\,.
\end{align}
Together with the minimum condition ($N_f=3$)
\begin{align}
    v = \frac{c +\sqrt{c^2+18 \lambda_\sigma m^2}}{3\lambda_\sigma}\,,
    \label{eq:vev-explicit}
\end{align}
which entails that a large $v$ corresponds to small $\lambda_\sigma$, we arrive at the desired conclusion that small $\lambda_\sigma$ corresponds to large $f_\pi$.

We infer the last key relation from \cref{eq:spectrum_chiral_limit}, which reads
\begin{align}
    m_\sigma^2&=\lambda_\sigma v^2-\left(1-\frac2{N_f}\right)m_{\eta}^2\,.
    \label{eq:msigma-explicit}
\end{align}
For fixed $m_{\eta}^2$, which is justified as explained before, $m_\sigma^2$ is directly related to $\lambda_\sigma v^2$. From \cref{eq:vev-explicit}, we know that a large $v$ corresponds to small $\lambda_\sigma$ but the explicit $\lambda_\sigma$ dependence in \cref{eq:msigma-explicit} is stronger and overall, we arrive at correlation that a larger $\lambda_\sigma$ corresponds to a larger $m_\sigma$, contrary to the naive statement in \cref{eq:msigma-tree}.

In summary, we have the following correlations:
\begin{itemize}
\item Smaller $m$ leads to smaller $m_\sigma$ and thus stronger GW signals, see the top left panel of \cref{fig:LSM_peak-coupling_binned}.
\item Smaller $\lambda_\sigma$ leads to smaller $m_\sigma$ and thus stronger GW signals. This is clearly shown in the top right panel of \cref{fig:LSM_peak-coupling_binned}.
\item Smaller $c$ leads to smaller $m_\sigma$ and thus stronger GW signals. See lower right panel of \cref{fig:LSM_peak-coupling_binned}.
\end{itemize}

These are the observations on the physical and theoretical parameters. From the physical nature, we observe the following interesting relations
\begin{itemize}
    \item \emph{Weak coupling regime}: Our model allows us to study the transition from an effective strong to an effective weak coupling regime. The strongest first-order phase transitions take place for small $\lambda_\sigma$ and $m^2$ couplings, where the latter is measured in units of $T_c$, see \cref{fig:LSM_peak-obs_binned}. Note, however, that this does not hold for the other couplings, $\lambda_a$ and $c$. This supports our previous findings \cite{Huang:2020crf, Reichert:2021cvs, Reichert:2022naa} where we observed that strongly coupled systems have a strongly suppressed GW signal. 
    \item \emph{Light sigma mass}: In the limit of a small sigma meson mass, the  state shares some similarities with the dilaton often associated with the dynamics of a near-conformal field theory. Near conformal theories are known to have a very strong first-order phase transition \cite{Witten:1980ez, Antipin:2012sm, Hambye:2013dgv, Sannino:2015wka, Iso:2017uuu, Chishtie:2020tze, Ellis:2020awk, Sagunski:2023ynd}, which we also observe here at small sigma masses.
    \item \emph{Supercooling}: The dominant reason for a strong first-order phase transition is typically a large temperature difference between the critical and the phase-transition temperature, also known as supercooling. We analysed this relation in \cref{fig:LSM_peak-cooling_binned} where we show the supercooling parameter $1-T_p/T_c$ as a function of the sigma mass $m_\sigma$ (left panel), the GW parameters $\tilde \beta$ (central panel) and $\alpha$ (right panel). We clearly observe the stronger phase transition with an increasing supercooling parameter. Note, the sharp edges for $N_f=3,4$ in the $\alpha$ plot (right panel), which are related to the edges already observed in \cref{fig:LSM_alpha-beta}. These edges limit the maximal strength of the phase transition since $\alpha$ increases very slowly despite a strongly increasing supercooling parameter. The exception is $N_f=5$ where $\alpha$ is continuously increasing with the supercooling parameter.
\end{itemize}

\section{Conclusions}
\label{sec:conclusion}

In this work, we explored the possibility that a strongly coupled dark sector may produce an observable stochastic gravitational wave signal. We employed the Polyakov Linear Sigma Model as an effective theory to study the first-order chiral phase transitions with $N_f=3,4,5$ flavours. We further implemented the well-established Cornwall-Jackiw-Tomboulis (known as CJT) method to bridge perturbative and non-perturbative regimes of the theory and thus are able to study both strongly- and weakly-coupled systems at once.

We observed that stronger first-order phase transitions and thus larger GW signals generically occur when the system features a light sigma meson $m_\sigma$ and/or is weakly coupled, corresponding to larger $f_\pi$.
The strongest phase transitions with $\tilde \beta = \mathcal{O}(10^3)$ and $\alpha = \mathcal{O}(10^{-1})$ are present for $N_f=3$ and small sigma mass. This is due to a zero temperature barrier in the potential induced by the cubic determinant term. It provides GW signals which are just outside the sensitivity of LISA, but that can be observable via the BBO and DECIGO detectors assuming that the confinement temperature is such that the peak frequency falls into the maximal sensitivity range of the detectors. We have also observed that the strength of the first-order phase transition generically increases with $N_f$ due to an increase of the latent heat (strength parameter $\alpha$).

The main features of our results are expected to be sufficiently general, robust and largely independent of the details of the model computations. This is so since all the physical underlying mechanisms at play have been embedded in the effective model description.

\begin{acknowledgments}
Z.-W.W.~thanks Guo Huai-Ke for helpful discussions. The work of F.S.~is partially supported by the Carlsberg Foundation, grant CF22-0922. M.R.~acknowledges support from the Science and Technology Research Council (STFC) under the Consolidated Grant ST/T00102X/1. R.P.~is supported in part by the Swedish Research Council grant, contract number 2016-05996, as well as by the European Research Council (ERC) under the European Union's Horizon 2020 research and innovation programme (grant agreement No 668679).
\end{acknowledgments}
 
\appendix

\begin{widetext}
\section{CJT computations for general $N_f$}
\label{sec:CJT_general_Nf}
In this appendix, we detail the derivation of the thermal masses within the CJT computation for general $N_f$. We follow \cite{Roder:2003uz} and also use their notation. The couplings are defined by
\begin{align}
\label{eq:coupling-redef}
\lambda_1 &= \frac12 (\lambda_\sigma - \lambda_a) \,,
&
\lambda_2 &= \frac{N_f}{2} \lambda_a \,.
\end{align}
Furthermore, we introduce the tensors $\mathcal F$, $\mathcal G$, and $\mathcal H$ describing the tensor structures in the potential, see (28) of \cite{Roder:2003uz}. We also use $\mathcal{S}$ for the full quantum propagator of the scalar particles ($\sigma$ and $a$) and $\mathcal{P}$ for the full quantum propagator of the pseudoscalar particles ($\pi$ and $\eta$). With these definitions, the relevant structures for the thermal masses are given by
\begin{align}
	4 \mathcal F _{00cd} \int\! \mathcal S_{cd} &= \frac{T^2}{2\pi^2} \left(\lambda_1 \left[3 I_B(R_\sigma) + (N_f^2-1) I_B(R_a) \right] + \frac{\lambda_2}{2} \left[ \frac6{N_f} I_B(R_\sigma) + \frac{6(N_f^2-1)}{N_f} I_B(R_a) \right]  \right), \notag \\
	4 \mathcal H _{00cd} \int\! \mathcal S_{cd} &= \frac{T^2}{2\pi^2} \left(\lambda_1 \left[ I_B(R_\sigma) + (N_f^2-1) I_B(R_a) \right] + \frac{\lambda_2}{2} \left[ \frac2{N_f} I_B(R_\sigma) + \frac{2(N_f^2-1)}{N_f} I_B(R_a) \right]  \right), \notag \\
	4 \mathcal F _{11cd} \int\! \mathcal S_{cd} &= \frac{T^2}{2\pi^2} \left(\lambda_1 \left[I_B(R_\sigma) + (N_f^2+1) I_B(R_a) \right] + \frac{\lambda_2}{2} \left[ \frac6{N_f} I_B(R_\sigma) + \frac{2( 2 N_f^2-3)}{N_f} I_B(R_a) \right]  \right), \notag \\
	4 \mathcal H _{11cd} \int\! \mathcal S_{cd} &= \frac{T^2}{2\pi^2} \left(\lambda_1 \left[ I_B(R_\sigma) + (N_f^2-1) I_B(R_a) \right] + \frac{\lambda_2}{2} \left[ \frac2{N_f} I_B(R_\sigma) +  \frac{2 (2	N_f^2-1)}{N_f} I_B(R_a) \right]  \right),
\end{align}
where $R_i\equiv M_i(\sigma,T)/T$ and the $I_B$ is given by \cref{eq:thermal-integrals}. The expressions are in straight analogy for the contractions with the pseudoscalars, $\int\! \mathcal P$. We furthermore have two contractions that are proportional to the determinant term, which are only relevant for the $N_f=4$ case
\begin{align}
	4 \mathcal G _{00cd} \int\! \mathcal S_{cd} &= - \frac{c}{2} \frac{T^2}{2\pi^2}  \left( \frac32 I_B(R_\sigma) - \frac{15}2 I_B(R_a) \right), \notag \\
4 \mathcal G _{11cd} \int\! \mathcal S_{cd} &= - \frac{c}{2} \frac{T^2}{2\pi^2}  \left( - \frac12 I_B(R_\sigma) + \frac{5}2 I_B(R_a) \right).
\end{align}
With these results and returning to our standard coupling conventions with \cref{eq:coupling-redef}, the thermal masses are given by
\begin{align}
    M_\sigma ^2 &= m_\sigma^2 + 4 [\mathcal F _{00cd} + \delta_{4N_f} \mathcal G _{00cd} ] \int\! \mathcal S_{cd}  + 4 [\mathcal H _{00cd} - \delta_{4N_f} \mathcal G _{00cd}] \int\! \mathcal P_{cd} \notag \\
    &= m_\sigma^2 + \frac{T^2}{4\pi^2} \bigg[ \left(3 \lambda_\sigma - \delta_{4N_f}  \frac32 c\right)  I_B(R_\sigma) + \left( (N_f^2-1) (\lambda_\sigma + 2\lambda_a) + \delta_{4N_f} \frac{15}2 c \right) I_B(R_a) \notag\\
    &\hspace{3cm}+ \left(\lambda_\sigma + \delta_{4N_f} \frac32 c\right) I_B(R_\eta) + \left((N_f^2-1)\lambda_\sigma - \delta_{4N_f} \frac{15}2 c \right) I_B(R_\pi)   \bigg], \notag \\
    M_a ^2 &= m_a^2 + 4 [\mathcal F _{11cd} + \delta_{4N_f}  \mathcal G _{11cd} ] \int\! \mathcal S_{cd}  + 4 [\mathcal H _{11cd} - \delta_{4N_f} \mathcal G _{11cd}] \int\! \mathcal P_{cd} \notag \\
    &= m_a^2 + \frac{T^2}{4\pi^2} \bigg[ \left(\lambda_\sigma + 2 \lambda_a + \delta_{4N_f} \frac12 c\right)  I_B(R_\sigma) + \left( (N_f^2+1) \lambda_\sigma + (N_f^2-4)\lambda_a - \delta_{4N_f} \frac{5}2 c \right) I_B(R_a) \notag \\
    &\hspace{3cm}+ \left(\lambda_\sigma - \delta_{4N_f}  \frac12 c\right) I_B(R_\eta) + \left((N_f^2+1) \lambda_\sigma + N_f^2\lambda_a + \delta_{4N_f}  \frac{5}2 c \right) I_B(R_\pi)   \bigg], \notag \\
    M_\eta^2 &= m_\eta^2 + 4 [\mathcal F _{00cd} + \delta_{4N_f} \mathcal G _{00cd} ] \int\! \mathcal P_{cd}  + 4 [\mathcal H _{00cd} - \delta_{4N_f} \mathcal G _{00cd}] \int\! \mathcal S_{cd} \notag \\
    &= m_\eta^2 + \frac{T^2}{4\pi^2} \bigg[ \left(3 \lambda_\sigma - \delta_{4N_f} \frac32 c\right)  I_B(R_\eta) + \left( (N_f^2-1) (\lambda_\sigma + 2\lambda_a) + \delta_{4N_f} \frac{15}2 c \right) I_B(R_\pi) \notag \\
    &\hspace{3cm} + \left(\lambda_\sigma + \delta_{4N_f} \frac32 c\right) I_B(R_\sigma) + \left((N_f^2-1)\lambda_\sigma - \delta_{4N_f} \frac{15}2 c \right) I_B(R_a)   \bigg], \notag \\
    M_\pi ^2 &= m_\pi^2 + 4 [\mathcal F _{11cd} + \delta_{4N_f} \mathcal G _{11cd} ] \int\! \mathcal P_{cd}  + 4 [\mathcal H _{11cd} - \delta_{4N_f} \mathcal G _{11cd}] \int\! \mathcal S_{cd} \notag \\
    &= m_\pi^2 + \frac{T^2}{4\pi^2} \bigg[ \left(\lambda_\sigma + 2 \lambda_a + \delta_{4N_f} \frac12 c\right)  I_B(R_\eta) + \left( (N_f^2+1) \lambda_\sigma + (N_f^2-4)\lambda_a - \delta_{4N_f} \frac{5}2 c \right) I_B(R_\pi) \notag \\
    &\hspace{3cm}+ \left(\lambda_\sigma - \delta_{4N_f} \frac12 c\right) I_B(R_\sigma) + \left((N_f^2+1)\lambda_\sigma + N_f^2\lambda_a + \delta_{4N_f} \frac{5}2 c \right) I_B(R_a)   \bigg].
\end{align}
These thermal masses are used in \cref{eq:VFT} and thereby provide the finite temperature part of the potential.

\end{widetext}

\bibliography{refs}

\end{document}